\documentstyle[aps,prb,twocolumn,eqsecnum,epsf,floats]{revtex}
\begin{document}

\draft
\preprint{}

\twocolumn[\hsize\textwidth\columnwidth\hsize\csname @twocolumnfalse\endcsname

\title{Magnetic-field-dependent zero-bias diffusive anomaly in Pb 
oxide-{\it n}-InAs structures: \\ Coexistence of two- and three-dimensional states}

\author{G. M. Minkov\cite{Minkov},
A. V. Germanenko, S. A. Negachev, O. E. Rut}
\address{Institute of Physics and Applied
Mathematics, \\ Ural
University, Ekaterinburg 620083, Russia}
\author{Eugene V.\ Sukhorukov\cite{Eugene}\cite{inst}
}
\address{Department of Physics and Astronomy,
         University of Basel,\\
         Klingelbergstrasse 82,
         CH--4056 Basel, Switzerland}
\date{\today}
\maketitle
\begin{abstract}
The results of experimental and theoretical 
studies of zero-bias anomaly (ZBA) in the Pb-oxide-n-InAs tunnel 
structures in magnetic field up to 6T are presented. 
A specific feature of the structures is a coexistence of the 2D and 3D 
states at the Fermi energy near the semiconductor surface. 
The dependence of the measured ZBA amplitude on the strength
and orientation of the applied magnetic field
is in agreement with the proposed theoretical model.
According to this model, electrons tunnel into 2D states, 
and move diffusively in the 2D layer, whereas the main contribution to the 
screening comes from 3D electrons.
\end{abstract}

\vskip2pc]
\narrowtext

\section{Introduction}
\label{sec:intro}

It is well known that the electron-electron interaction
strongly influences the transport properties of disordered 
conductors.\cite{Altshu85} Even in the presence of 
weak disorder ($\varepsilon_{F}\tau_p \gg 1$, 
where $\varepsilon_{F}$ is the Fermi energy, $\tau_p$ 
is the momentum relaxation time, and $\hbar = 1$)
the electron-electron interaction suppresses the one-particle 
density of states at the Fermi level (diffusive anomaly). 
This leads to  small deviations from Ohm's law in the current-voltage 
characteristics of a tunnel junction at small voltages $V$. 
The diffusive anomaly, which appears as a dip in the differential tunneling 
conductance $G=dI/dV$ at zero bias, 
reveals itself in almost all tunneling experiments 
and has been studied in various tunneling structures.\cite{Wolf} 
This should be distinguished from other nonlinearities of the current-voltage 
characteristics at low bias, which are due to 
different physical phenomena.
The form of the diffusive zero-bias anomaly (ZBA) depends on the 
dimensionality: $\delta G(V)\propto \ln |V|$ for
tunneling into two-dimensional (2D) conductors, and 
$\delta G(V)\propto \sqrt{|V|}$
for three-dimensional (3D) conductors. The width of the dip
in the tunneling conductance is of order $\tau_p^{-1}$, and therefore 
cannot be observed in pure conductors.

The first theoretical explanation of the diffusive ZBA,
by Altshuler, Aronov and Lee in Refs.\ \onlinecite{Lee}
and \onlinecite{Altshu79},
was based on the diagrammatic perturbative method. 
For low-dimensional systems,
this theory was subsequently extended beyond the perturbative
treatment by Nazarov in
Refs.\ \onlinecite{Nazarov1}
and
\onlinecite{Nazarov2} (see also Refs.\ \onlinecite{Devoret,Ingold},
where the realistic system is described by the coupling of the 
tunnel junction with the effective electromagnetic environment), 
and later, by Levitov and 
Shytov in Ref.\ \onlinecite{Lev}. 
Nazarov also gave a transparent physical interpretation 
of the diffusive ZBA;
immediately after an electron 
tunnels into the diffusive conductor and forms the distribution 
$\rho ({\bf r},t)$, the system acquires an extra 
energy due to the interaction
between this electron and the electrons in the conductor
(Coulomb barrier).
Therefore, the electron density
perturbation $\rho ({\bf r},t)$ must spread under the Coulomb barrier 
in order to reach the final state.
This process contributes a many-electron action $S(t)$
($t\sim 1/eV$ is the time of spreading of the electron density
perturbation)
to the total tunneling action, and thereby, suppresses the
tunneling current. In the regime of the Coulomb blockade effect
($S(t)\gg 1$) the tunneling current is almost completely
suppressed. Conversely, 
in good metals ($\varepsilon_{F}\tau_p \gg 1$)
the Coulomb interaction is screened, so that the many-electron action
is small, $S(t)\ll 1$, and gives only small correction to
the differential conductance. At $T=0$, this takes the form
\begin{equation}
{{1}\over{G}}{{d G}\over {d V}}={{2e}\over {\pi}}Im\left\{\left
.S(\omega )\right|_{i\omega\to -eV+i0}\right\}\;.
\label{IV}
\end{equation}
The  density of the tunneling electron $\rho_{\omega}({\bf r})$
is given by a diffusion propagator (diffuson),
whereas the electrodynamical potential $\phi_{\omega}({\bf r})$ 
which it excites is     given by
\begin{equation}
\phi_{\omega}({\bf r})=\int d{\bf r}^{\prime}V_{\omega}({\bf r},
{\bf r}^{\prime})\rho_{\omega}({\bf r}^{\prime}),
\label{phi}
\end{equation}
where $V_{\omega}({\bf r},{\bf r}^{\prime})$ is the dynamically
screened Coulomb potential. The action $S(\omega )$ is then explicitly
given by: \cite{Nazarov1}
\begin{equation}
S(\omega )={1\over 2}\int d{\bf r}\;\rho_{-\omega}({\bf r})\phi_{
\omega}({\bf r}).
\label{action}
\end{equation}

This simple formula for the action displays an important role
for the interface of the tunnel junction in the ZBA in the case of tunneling
into a 3D conductor. 
Indeed, after the electron tunnels
through the barrier, it first appears on the surface of the conductor
before propagating into the bulk. The surface of the conductor obviously
affects the spreading process of the electron density $\rho_{\omega}$.
Consequently, it affects the amplitude of the ZBA. For example,
it can partially block the spreading of the electron into final state,
giving rise to additional factor of $2$
 in the amplitude of the anomaly \cite{Altshu84}
(the electron propagates into the half space).
This interface effect is even more pronounced in the presence of a 
magnetic field.
The role of the magnetic field is twofold. It causes the Lorentz force,
which blocks the spreading of the electron density, but it also
induces a Hall
voltage, which causes a drift along the interface,
and thereby enhances the spreading.
If the magnetic field ${\bf B}$ is perpendicular to the junction interface,
only the first effect contributes to the ZBA and gives a $B^2$ dependence
of the ZBA.\cite{Altshu79}
If the magnetic field is parallel to the junction interface, the two effects
exactly cancel. 
This results in the strongly anisotropic magnetic field 
dependence of the ZBA predicted in Ref.\ \onlinecite{Sukhor}.
Namely, the ZBA depends only on the component
of the magnetic field perpendicular to the interface of the junction, as 
it would be in the case of tunneling into a 2D conductor. This effect has
probably been observed in Refs.\ \onlinecite{anisotropy}
and \onlinecite{Dub}. 

Motivated by this physical situation, we theoretically and experimentally
investigated the ZBA  in Pb-oxide-n-InAs 
structures in the presence of a magnetic field.
We expected that the specific feature of 
these structures, namely, coexistence of 3D and 2D electron states near the 
surface of InAs, will strongly influence the ZBA and especially its magnetic field
dependence. In particular, as the current in these structures can occur through 
the tunneling  of electrons into both 2D and 3D states, the principle question
which arises is whether the ZBA has 2D or 3D character. The results of our study can be 
summarized as follows. The electrons tunnel into 
2D states and move diffusively in a 2D layer, whereas the main contribution 
to the screening comes from 3D electrons. This gives rise to the unusual
magnetic field dependence of the ZBA. 
When the magnetic field {\bf B} is perpendicular to the interface of the tunnel 
junction, the amplitude of the ZBA grows as $B^2$  
in agreement with Ref.\ \onlinecite{Altshu79}.
The ZBA amplitude strongly depends on the orientation of the 
magnetic field, in agreement with Ref.\ \onlinecite{Sukhor}.
However, when the magnetic field lies in the plane of the junction interface, 
the magnetic field dependence does not disappear.
Instead, the ZBA amplitude is linear in $B$.
%
\begin{figure}[p]
  \begin{center}
    \leavevmode
\epsfxsize=7.5cm
 \epsfclipon
\epsffile{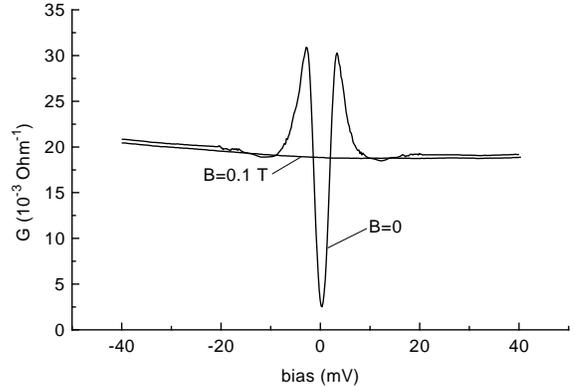} 
  \end{center}
\caption{
Bias dependencies of the differential conductance $G$ 
for structure 1 at T= 1.6 K.}
\label{fig1}
\end{figure}
%
%
\begin{figure}[p]
  \begin{center}
    \leavevmode
\epsfxsize=8.0cm
 \epsfclipon
\epsffile{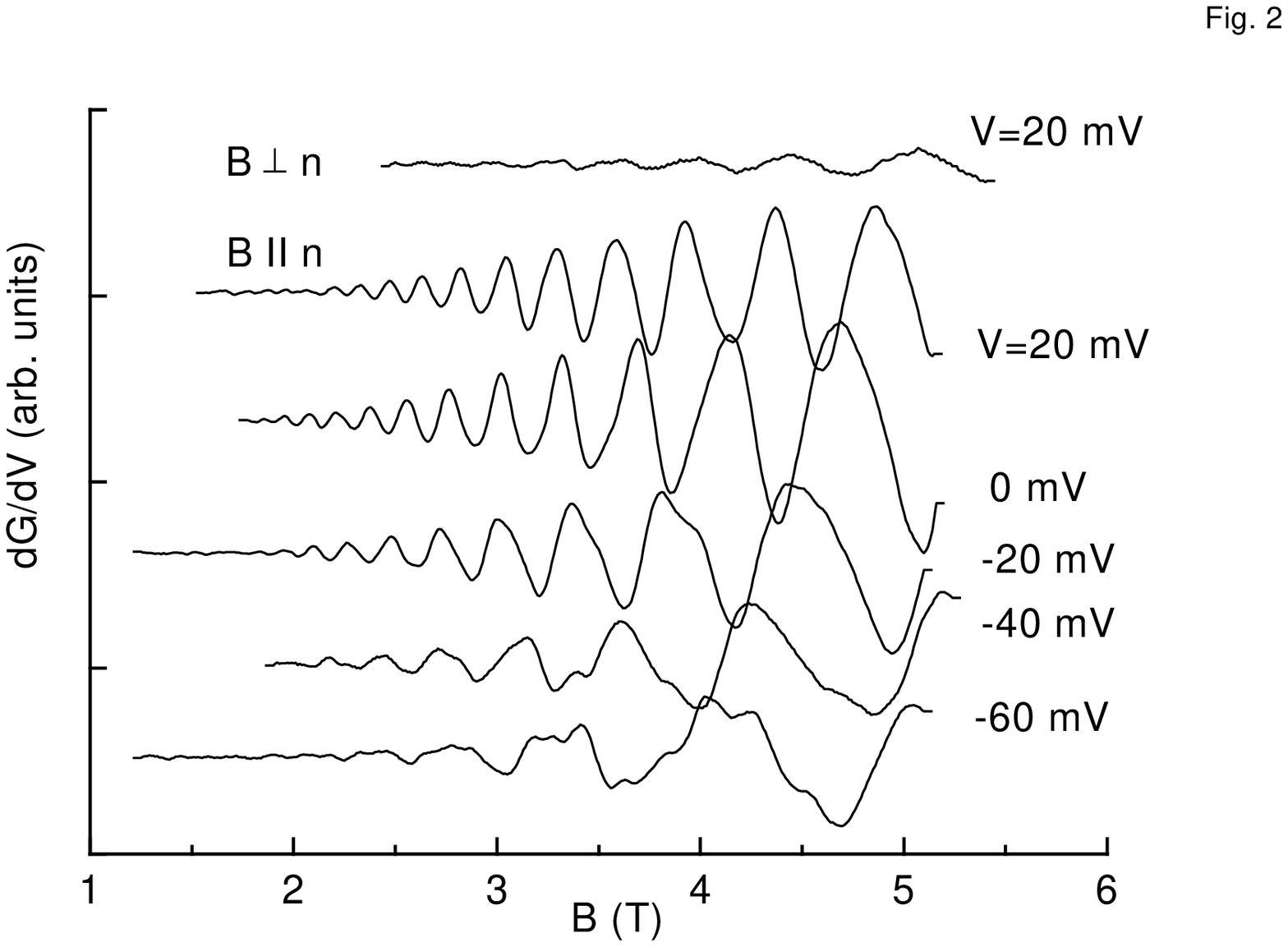} 
  \end{center}
\caption{Magnetic field dependence of  $dG/dV$ for different
biases. The topmost curve is for ${\bf B}\perp {\bf n}$, and the others 
are for
${\bf B}\parallel {\bf n}$ (structure 1). All curves are at T= 4.2 K.}
\label{fig2}
\end{figure}
\begin{figure}[p]
  \begin{center}
    \leavevmode
\epsfxsize=8.0cm
\epsfclipon
\epsffile{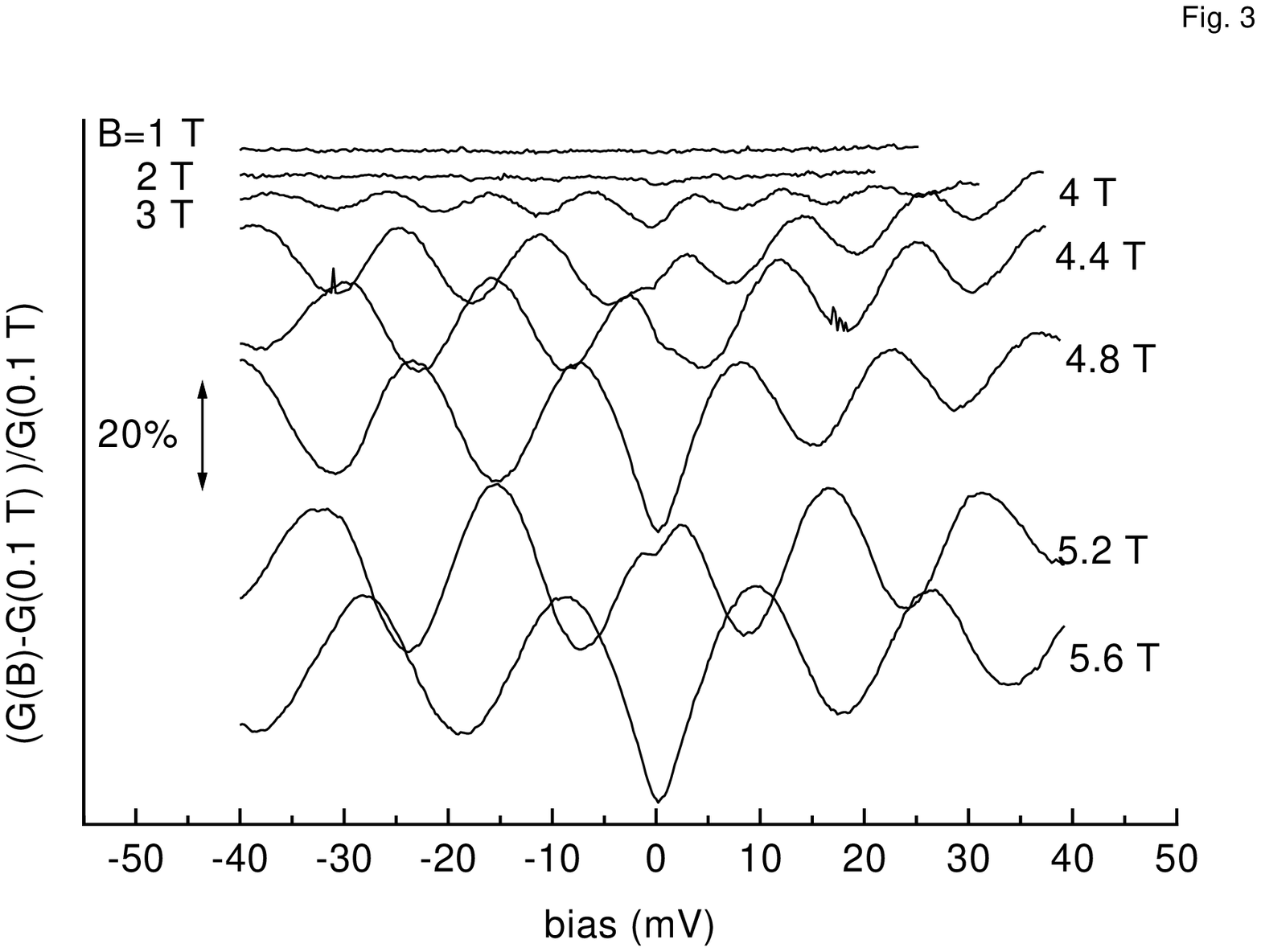} 
  \end{center}
\caption{Bias dependencies of $(G(B)-G(0.1 T))/G(V,0.1 T)$ for 
different magnetic fields
${\bf B}\parallel {\bf n}$ at T= 4.2 K.}
\label{fig3}
\end{figure}

\section{Experimental results }

The differential conductance  $G=dI/dV$  and its
derivative $(dG/dV)$ as a function of bias and magnetic field in
Pb-oxide-n-InAs tunnel structures were investigated in a magnetic
field up to 6 T  at temperatures 4.2 and 1.6  K.
The tunnel
structures were fabricated on  n-InAs wafers with 
two different pairs of electron concentration  and mobility:
$9.7\times 10^{17}$ $cm^{-3}$, and
$1.5\times 10^4$ $cm^2V^{-1}s^{-1}$ (structure 1);  
and $1.9\times 10^{17}$ $cm^{-3}$, and 
$1.8\times 10^4$ $cm^2V^{-1}s^{-1}$ (structure 2).  Ultraviolet
illumination for 10--15 min in dry air was used to form the thin
oxide, which served as a tunneling barrier. The Pb electrode
was then evaporated through a mask. The tunnel contacts fabricated on each wafer
were similar and results are shown for one of several contacts
fabricated on each wafer.  The
traditional modulation procedure was used for
measuring the differential conductance and its derivative.
Measurements showed that decreasing of the
modulation amplitudes below 0.2 mV do not change the
features in the $G$ {\it vs.}\ $V$ curves. Therefore, in all investigations
the modulation amplitude was 0.2 mV.

The dominant contribution to the  current in the investigated structure is a 
tunneling current. This is evident from  the bias dependencies of the 
differential conductance, which are shown on Fig.\ \ref{fig1}. 
The structure of
the curve for $B=0$  is the ``superconducting anomaly'', connected
with the superconducting gap in the one particle density of states in the metal 
electrode. 
At $B>0.06 $ T the superconductivity of Pb is destroyed and
this structure disappears completely. 

\begin{figure}[p]
  \begin{center}
    \leavevmode
\epsfxsize=5.5cm 
 \epsfclipon
\epsffile{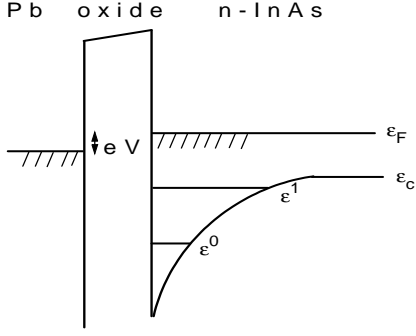} 
  \end{center}
\caption{Energy diagram of the Pb-oxide-n-InAs  tunnel structure. 
$\varepsilon^0$ and $\varepsilon^1$ are the energies of the bottom 
of the ground and excited 2D subbands respectively, and 
$\varepsilon_c$ is the energy of the bottom of the conduction band.}
\label{fig4}
\end{figure}

Oscillations 
in $G$ and $dG/dV$ as a function of $V$ and $B$
were observed for both
${\bf B}\parallel {\bf n}$ and ${\bf B}\perp
{\bf n}$, where ${\bf n}$ is  the
normal to the plane of the tunnel junction (Figs.\ \ref{fig2} and \ref{fig3}). 
The tunneling conductance oscillations in such types of structures
were comprehensively studied in InAs,\cite{Tsui1,Tsui2,Tsui3} and in 
HgCdTe. \cite{Minkov95,Minkov73} 
It was shown that in the structures based on InAs, an
accumulation layer with 2D subbands exists near the barrier
(Fig.\ \ref{fig4}). The tunneling conductance is determined by 
tunneling into both  3D and  2D states of the semiconductor electrode. 
Having ${\bf B}\parallel {\bf n}$ leads to quantization of the
spectrum of both 2D and 3D states. For this orientation
of the magnetic field, the oscillations in $G$
are mainly due to the modulation of the density of 2D states. 
Having ${\bf B}\perp {\bf n}$  does not
quantize the energy spectrum of 2D states and the oscillations in 
$G$ are only due to tunneling into 3D states. At fixed bias $V$,
these
oscillations are periodic in $1/B$.
Therefore,  using the Fourier
transformation one can determine the fundamental fields $B_f$ and,
consequently, the quasi-momenta $k =\sqrt{2eB_f/c\hbar}$
of 2D and bulk states at the energy $\varepsilon_F+eV$. 
In addition, such data processing allows us to determine the energies of the 
bottoms of the conduction band and 2D subbands counted from the Fermi energy
of the semiconductor (for more details, see Refs.\ \onlinecite{Tsui1} 
and \onlinecite{Minkov95}).  
Thus, we found that in structure 1 there are bulk states with
$\varepsilon_F-\varepsilon_c=115$ $meV$ 
and $k_b^2(\varepsilon_F)=9.3\times 10^{12}$ $cm^{-2}$, 
states of the ground 2D subband with
$\varepsilon_F-\varepsilon^0\simeq 160$ $meV$ 
and $k^2_0(\varepsilon_F)=20.6\times 10^{12}$ $cm^{-2}$,  
and states of the excited 2D subband with
$\varepsilon_F-\varepsilon^1\simeq 120$ $meV$ and
$k_1^2(\varepsilon_F)=10.3 \times 10^{12}$ $cm^{-2}$.
For the structure 2 these parameters are 
$\varepsilon_F-\varepsilon_c=50$ $meV$ and 
$k^2_b(\varepsilon_F)=3.1\times 10^{12}$, 
$\varepsilon_F-\varepsilon^0\simeq 95$ $meV$
and $k^2_0(\varepsilon_F)=7.6\times 10^{12}$, 
and $\varepsilon_F-\varepsilon^1\simeq 55$ $meV$ 
and $k^2_1(\varepsilon_F)=3.5 \times 10^{12}$ $cm^{-2}$.
%
%
\begin{figure}[p]
  \begin{center}
    \leavevmode
\epsfxsize=8.0cm
 \epsfclipon
\epsffile{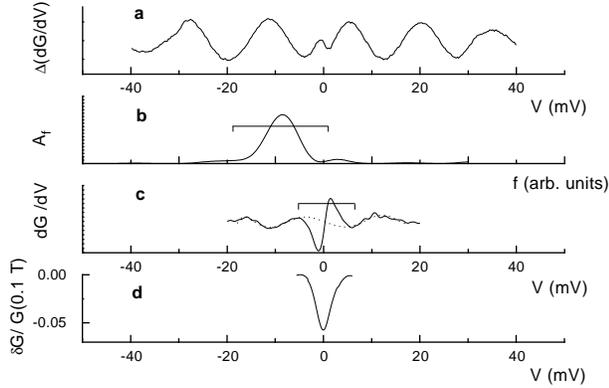} 
  \end{center}
\caption{(a) The bias dependence of 
$\Delta(dG(V,B)/dV)=$ $dG(V,B)/dV-dG(V,0.1\text{ T})/dV$ 
for $B=5.3$ T and ${\bf B}\parallel {\bf n}$. 
(b) Fourier transform of the upper curve. The bar shows the region 
which was cut out. (c) Result of inverse Fourier
transform. The dotted curve is the interpolation after 
the central region was cut out. (d) Reconstructed zero-bias anomaly
after the processing described in the text. }
\label{fig6}
\end{figure}
\begin{figure}[p]
  \begin{center}
    \leavevmode
\epsfxsize=8.0cm
 \epsfclipon
\epsffile{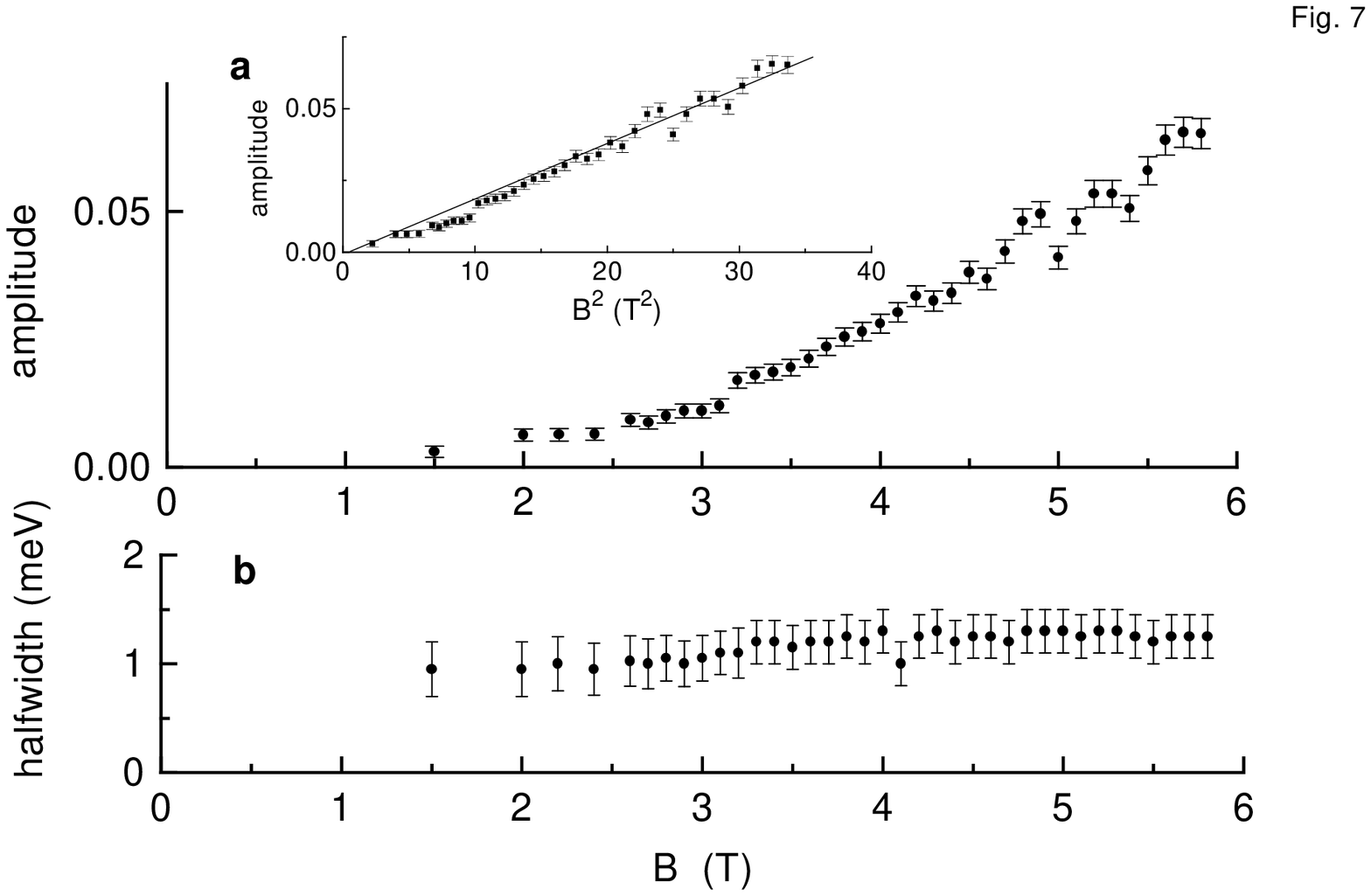} 
  \end{center}
\caption{Magnetic field dependencies of the amplitude (a) and 
halfwidth (b) of the ZBA, for ${\bf B}\parallel {\bf n}$. The inset shows the
$A$ {\it vs.}\ $B^2$ dependence. All points are at T= 4.2 K.}
\label{fig7}
\end{figure}
%

Now let us consider $G$ {\it vs.}\ $V$ curves in
the vicinity of zero bias. The relative difference 
$(G(V,B)-G(V,0.1\text{ T}))/$ $G(V,0.1\text{ T})$  
as a function of  voltage for various magnetic fields 
${\bf B}\parallel {\bf n}$  is presented in Fig.\ \ref{fig3}. 
It is seen that increasing $B$ gives rise to a 
dip in the conductance  in the vicinity of $V=0$, which is
better seen when it falls between adjacent 2D Landau levels.
This peculiarity is more pronounced in
$\Delta\left(dG(V,B)/dV\right)=dG(V,B)/dV-dG(V,0.1\text{ T})/dV$
{\it vs.}\  $V$ curves 
(Fig.\ \ref{fig6}a). 
To separate out the
ZBA from the conductance oscillations (due to 
Landau quantization), the following procedure was used; after
taking the Fourier transformation (Fig.\ \ref{fig6}b) we cut out the
components associated with the oscillations and then take the
inverse Fourier transformation (Fig.\ \ref{fig6}c). 
Such a procedure greatly helps in extracting the anomaly from the oscillations,
but does not completely separate the ZBA from the oscillations.
Therefore, we cut out the part of the curve in
the range $\pm 5$ mV in vicinity of $V=0$  (Fig.\ \ref{fig6}c), 
interpolate the rest of the curve by a smooth line, and 
then subtract this line
from the initial curve shown in Fig.\ \ref{fig6}c.  
After integration, we obtain the ZBA
in the tunneling conductance (Fig.\ \ref{fig6}d). (The
correctness of such processing was verified by separating out
the Gaussian shape from the simulating curve
$A_1\sin(\omega_1 V+\varphi_1)+A_2\sin(\omega_2 V+\varphi_2)+ A_3 \exp(-
(V/\Delta)^2$.)

\begin{figure}[h]
  \begin{center}
    \leavevmode
\epsfxsize=7.5cm
 \epsfclipon
\epsffile{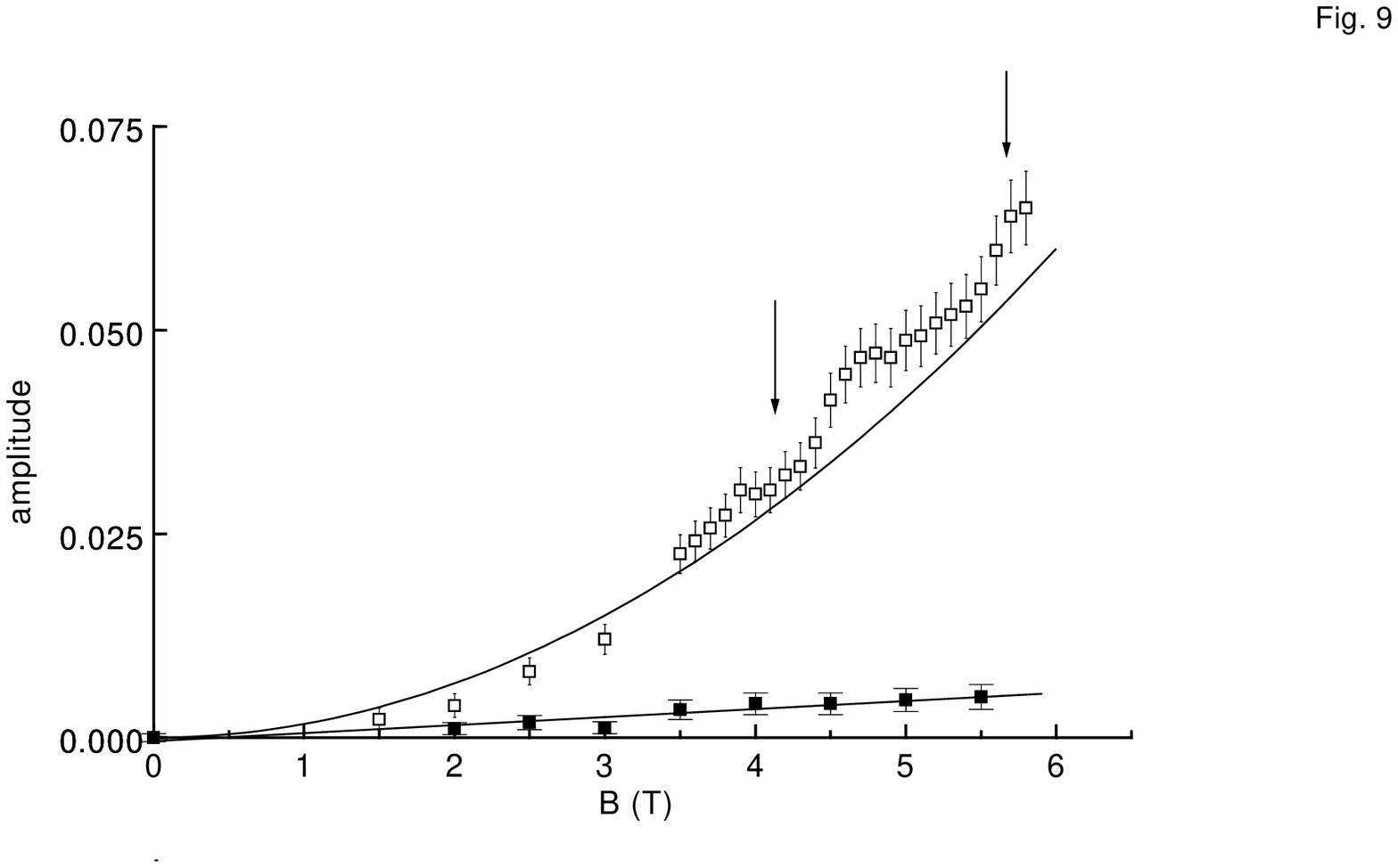} 
  \end{center}
\caption{Magnetic field dependencies of the ZBA amplitude  
for ${\bf B}\parallel {\bf n}$
(open squares) and ${\bf B}\perp {\bf n}$ (full squares) 
for structure 2 at T= 4.2 K. The arrows indicate the magnetic fields, 
for which the  
2D Landau levels coincide with the Fermi level.}
\label{fig9}
\end{figure}

The magnetic field dependencies  of the normalized amplitude
$A=-\delta G/G|_{V=0}$ 
and halfwidth of the ZBA  
are plotted in Fig.\ \ref{fig7}. 
It is seen that the halfwidth does not vary with the magnetic field  
within the experimental error,  whereas the amplitude of the ZBA significantly 
increases. The inset in Fig.\ \ref{fig7}a shows that the $A$ {\it vs.}\ $B$ 
dependence is close to $A\propto B^2$. 
Similar results are obtained for the structure 2 (Fig.\ \ref{fig9}).
In addition, one can see an oscillatory dependence of $A$ on $B$,
which appears at high magnetic fields.
The minima of the oscillations are observed 
at those magnetic fields where the 2D Landau 
levels cross the Fermi level. 
Thus, the origin of the oscillations of the ZBA amplitude
is the Landau quantization of electron states in the 2D layer.
The detailed investigation of this effect will be the subject of future work. 
Therefore, we will not concentrate on these oscillations in this paper.
%
%
%
\begin{figure}[h]
  \begin{center}
    \leavevmode
\epsfxsize=7.5cm
 \epsfclipon
\epsffile{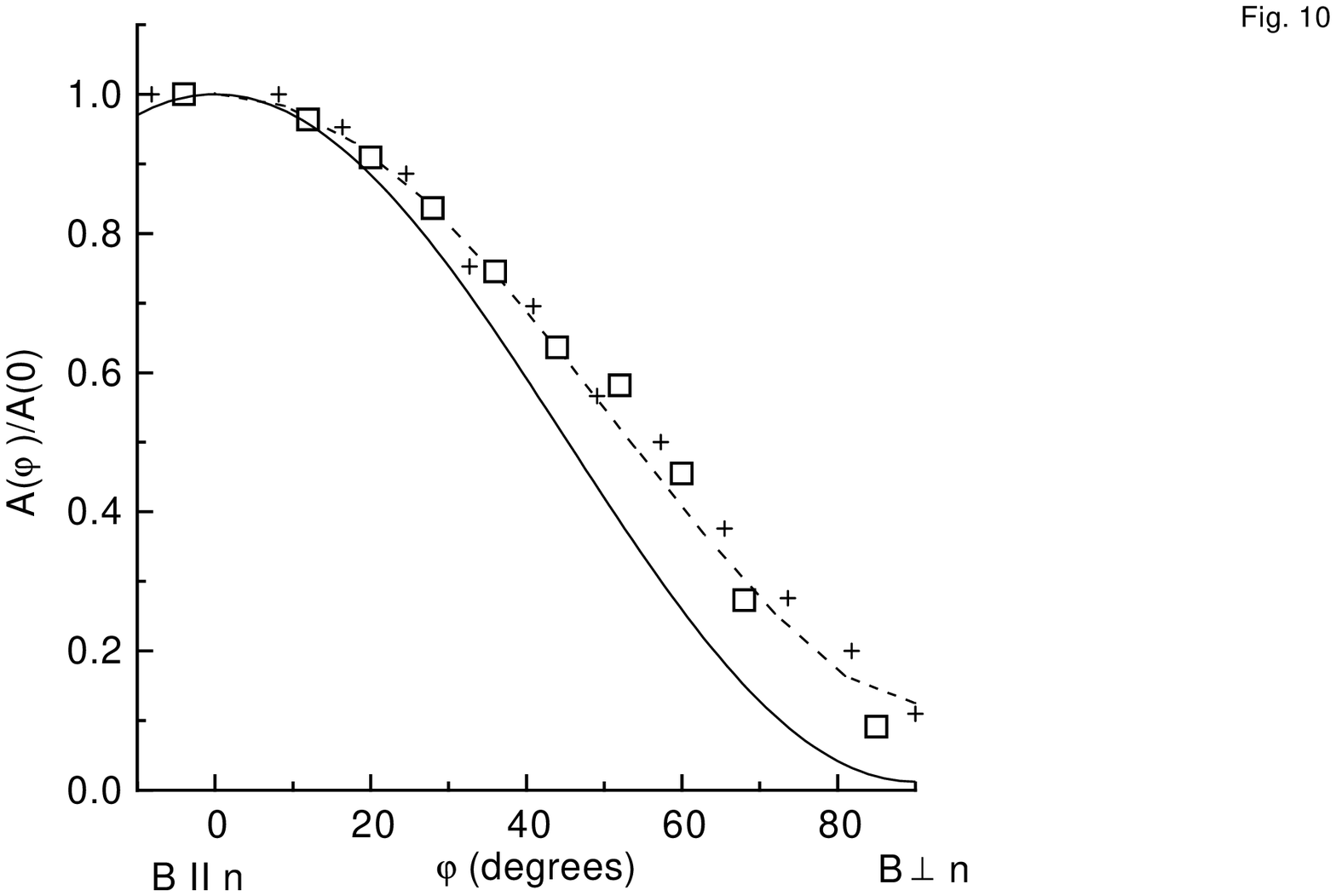} 
  \end{center}
\caption{Angular dependence of the ZBA amplitude, for B=5.5 T. 
The crosses and squares are data for structures 1 and 2 respectively.  
The dashed curve is the result of the calculation described in the
text. The solid curve corresponds to the 3D case.}
\label{fig10}
\end{figure}
%
\begin{figure}[t]
  \begin{center}
    \leavevmode
\epsfxsize=8.0cm
 \epsfclipon
\epsffile{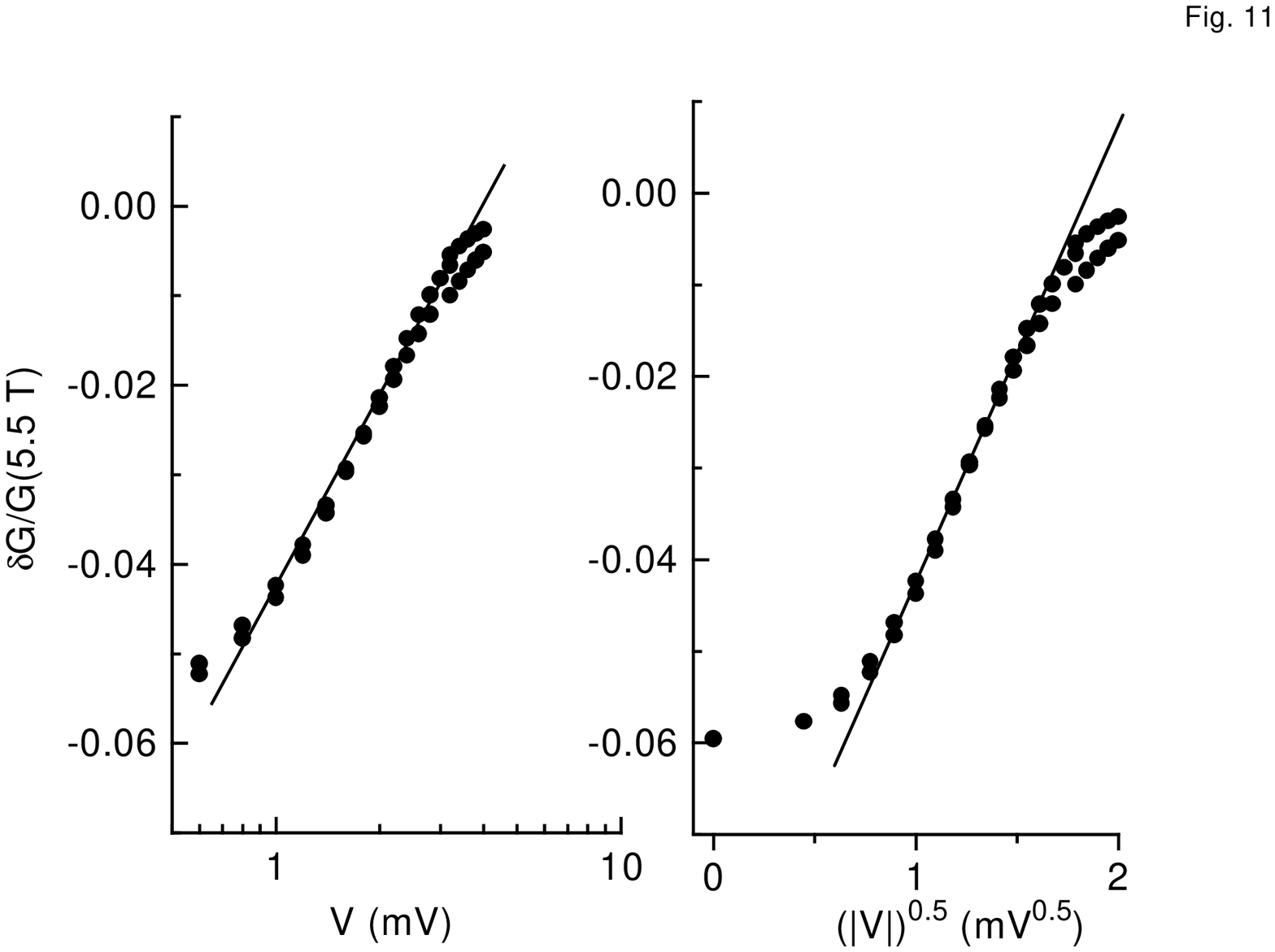} 
  \end{center}
\caption{Tunneling conductance near $V=0$ versus
$\ln |V|$ (a) and versus $|V|^{0.5}$ (b) at T= 1.6 K, B=5.5 T 
The two sets of data points correspond to different signs of the bias.}
\label{fig11}
\end{figure}

The angular dependence of the ZBA amplitude is plotted in
Fig.\ \ref{fig10} ($\varphi$ is the angle between ${\bf B}$ and ${\bf n}$). 
One can see that the ZBA amplitude is strongly anisotropic. 
It drastically decreases when the magnetic field deviates from 
${\bf B}\parallel {\bf n}$, but it does not disappear at 
${\bf B}\perp {\bf n}$.
The magnetic field dependence of the ZBA amplitude
for this orientation is significantly weaker and 
is close to linear (Fig.\ \ref{fig9}).

\section{Discussion}
\label{sec:disc}

The energy scale of the diffusive anomaly caused by electron-electron 
interaction in dirty conductors is  $\tau_p^{-1}$. 
In our structures, 
$\tau_p^{-1}$ is estimated from 
the mobility to be about 2 meV, whereas the halfwidth of the ZBA is 1 meV 
(Fig.\ \ref{fig7}b). Thus, we suppose that the ZBA observed in our 
experiment is just the diffusive anomaly.
The specific feature of the investigated structures is the coexistence 
of 2D and 3D electrons near the barrier. Therefore the basic question 
is whether the zero-bias anomaly is due to the interaction 
of 2D or 3D electrons.
In principle, the ZBA has different form for tunneling into 2D and 3D
states: $\delta G(V)\propto ln|V|$ for 2D and 
$\delta G(V)\propto\sqrt {|V|}$ for 3D.
However, the comparison of fits to experimental 
data in Fig.\ \ref{fig11} does not allow us to distinguish between the
two forms of the ZBA.

On one hand, the main part of the tunneling conductance 
is due to tunneling  into the 2D states. This  follows from the theoretical 
calculation of the tunneling conductance  
for the investigated structures carried out in the framework of the
transfer Hamiltonian method. \cite{Bardin}
Such a calculation shows that the tunneling conductance due to tunneling 
into the 2D states is larger by about a factor 
of 5 than that for tunneling into 3D states. 
This conclusion is also supported by the fact that  the amplitude of
oscillations of the tunneling conductance caused by the Landau quantization  
is significantly larger in the case of tunneling into 2D states 
(at ${\bf B}\parallel {\bf n}$) than in the case of tunneling into 
3D states (${\bf B}\perp {\bf n}$) (Fig.\ \ref{fig2}).
In addition, the ZBA amplitude has typical for 2D systems strong 
dependence on the magnetic field orientation, 
i.e.\ it is determined mainly by the normal component of the 
magnetic field (Fig.\ \ref{fig10}).
Therefore, one can surmise that the ZBA has 2D character.

On the other hand, the strong angular dependence of the 
ZBA is ambiguous evidence of the 2D nature of the ZBA.
Indeed, in Ref.\ \onlinecite{Sukhor} it was 
demonstrated that for tunneling into
3D states the amplitude of the ZBA is given by 
$A(B)\propto 1+\omega_c^2\tau_p^2\cos^2(\varphi )$ 
(where $\omega_c$ is cyclotron frequency), i.e.\ 
it depends only on the component
of the magnetic field perpendicular to the interface of the junction, as 
it would be in the case of tunneling into a 2D conductor. 
In the limit $\omega_c^2\tau_p^2\gg 1$  this leads to the strong angular 
dependence of the ZBA amplitude. Although the strong anisotropy of the ZBA 
is observed in our experiment for $\omega_c^2\tau_p^2\simeq 50$ at $B=5.5$ T,
the curve $1+\omega_c^2\tau_p^2\cos^2(\varphi )$
does not fit well with the experimental data (see the Fig.\ \ref{fig10}).
Moreover, the ZBA amplitude is linear in $B$ when the magnetic field 
lies in the plane of the junction interface (see Fig.\ \ref{fig7}).

Thus, the magnetic field dependence of the ZBA 
in the investigated structures 
does not completely agree with either the 
3D or 2D nature of the ZBA. 
We would like to stress however, that our experimental set up is 
not usual for studying the diffusive ZBA. Traditionally,
the 2D metallic layer in tunnel junctions is electrically isolated
from the 3D electrode (or another 2D layer).
In this case the charge relaxation is two-dimensional and 
the interaction is partially screened by the 3D metal, so that its  
strength is defined by the distance $\Delta$ between 2D 
and 3D electrodes. The correction to the differential conductance
then has the form\cite{Altshu84}
$\delta G(V)\sim\ln(a\Delta/r_{D}^2)\ln(eV\tau_p)$,
where $a$ is the width of the 2D layer, and $r_{D}$ is the Debye radius.
In addition, it is assumed that this formula holds for
$a\Delta/r_{D}^2\gg 1$ and $\Delta/r_{D}\gg 1$. Thus, there are two reasons
which make our experimental set up different from the usual one, and the
above formula nonapplicable to our case.
The specific feature of the investigated structures is  
coexistence of 3D and 2D electron states near the surface of the semiconductor
(see Fig.\ \ref{fig4}).
Thus, formally in our case $\Delta = 0$. Second, 
and this is most important, in our experiment the 2D electron
system and 3D metal are not electrically isolated.

In the next section we show that
the following scenario of tunneling is realized in this structures.
After tunneling, the electron moves diffusively in the 2D layer and forms 
the 2D distribution $\rho_{\omega}({\bf r})$ at the surface 
of the semiconductor. It immediately pushes other electrons into the 
bulk, so that the total charge becomes zero for a short time of order
of the inverse plasma frequency. Thus, the relaxation of the
{\em total} charge 
takes three-dimensional form, and this should lead to $\sqrt{V}$-dependence 
of the differential conductance usual for the 3D ZBA.
The dimensionality of  $\rho_{\omega}({\bf r})$ does not 
affect the voltage dependence of the differential conductance.
However, it leads to the unusual magnetic field dependence of the ZBA 
discussed above.

\section {Theoretical model and comparison with experimental data }
\label{sec:theor}

In our theoretical analysis we make two assumptions.
First, we assume that the probability of electrons tunneling 
into 2D states near the 
surface of the semiconductor is much greater than the probability
of tunneling into 3D bulk states. 
This  follows from the analysis of the experimental results in the 
previous sections. 
The second assumption is that the tunneling electron (though being 
screened
by the other electrons of the system) 
remains in the 2D well for a physically 
relevant time, i.e.\ for the time $t\sim 1/eV$ 
(see discussion in Sec.\ \ref{sec:intro})
before escaping into the bulk. This last time ranges from $\tau_p$
to $(kT)^{-1}$ in the ZBA regime, 
and thus in our experiment it may exceed the momentum 
relaxation time by factor of 10.
On the other hand, the escape of electrons from
the 2D well to the bulk is determined by
the ionized impurity scattering, which is main scattering mechanism at low
temperatures. This mechanism is strongly anisotropic --
the small angle scattering dominates.
Together with requirement of large  momentum 
transfer for  2D$\to$3D transition this leads to the fact that 
the  2D $\to$ 2D transition rate $W_{22}\sim (\tau_p)^{-1}$  is
larger than the 2D $\to$ 3D transition rate, $W_{23}$.
The calculations with wave functions and screening 
radius corresponding to the investigated structures carried out in the same 
manner as in Ref.\ {\onlinecite{rate23}}
gives $W_{22}/W_{23}\approx 15$.
Thus, the second assumption is justified.

The width of the 2D well is of order $\lambda_F$ and is much smaller 
than the mean free
path $l_3$ in the bulk of the conductor. Therefore, the 2D well
can be thought of as a $\delta$-layer with respect to 
the physically relevant length
scale. This means that after tunneling the electron forms a 2D density
distribution $\rho_{\omega}({\bf r})=Q_{\omega}({\bf R})\delta (z
)$ localized on the surface $z=0$ of
the conductor (here, ${\bf R}=(x,y)$ is the coordinate on the surface,
and ${\bf r}=({\bf R},z)$).
Then, due to the interaction of this electron with the ones forming 
both 2D and 3D liquids, the electrodynamic potential $\phi_{\omega}$ is excited.

Instead of a direct calculation of the integral (\ref{phi}) for
$\phi_{\omega}$,
we follow Ref.\ \onlinecite{Sukhor} and use the electroneutrality principle.
We assume that the density of the tunneling electron is
completely screened on the distance of the order of Debye radius $r_D$,
so that the induced charge density is $\tilde{\rho}_{\omega}({\bf r}
)=-\rho_{\omega}({\bf r})=-Q_{\omega}({\bf R})\delta (z)$.
Taking into account charging effects \cite{Sukhor} gives only corrections
of order $r_D^2/wl_3\ll 1$ ($w$ is the thickness of the tunneling barrier), 
which we neglect here. We also assume
that the Pb electrode, being a good metal, does not contribute to
the action $S(\omega)$. Therefore, after Fourier transformation
the integral (\ref{action}) can be represented in the following form:
\begin{equation}
S(\omega )={1\over {8\pi^2}}\int d{\bf k}\;Q_{-\omega}(-{\bf k})
\Phi_{\omega}({\bf k}),
\label{action2}
\end{equation}
where $\Phi_{\omega}({\bf k})=\phi_{\omega}({\bf k},z)|_{z=0}$.

The density $Q_{\omega}$ obeys the 2D diffusion equation in imaginary time
\begin{equation}
|\omega |Q_{\omega}-D_2\nabla^2_{{\bf R}}Q_{\omega}=
-e\:sign(\omega )
\delta ({\bf R}-{\bf R}_0),
\label{diffeq}
\end{equation}
with the diffusion coefficient $D_2\equiv D_2({\bf B})$, which
depends only on the $z$-component of the magnetic field 
${\bf B}=(B,\varphi )$:
\begin{equation}
D_2({\bf B})={{D_2^{\scriptscriptstyle (0)}}
\over {1+\omega_c^2\tau_2^2\cos^2\varphi}}.
\label{D2}
\end{equation}
Here, $\omega_c$ is the cyclotron frequency
and $D_2^{\scriptscriptstyle (0)}\equiv D_2(0)$.
We have introduced the new notation, $\tau_2$, for the momentum relaxation time
of 2D electrons to distinguish it from that of 3D electrons. 
After Fourier transformation,
equation (\ref{diffeq}) can be immediately solved,
\begin{equation}
Q_{\omega}({\bf k})=-{{e\:sign (\omega )}\over {|\omega |+D_2
{\bf k}^2}}.
\label{diffuson}
\end{equation}

To calculate the potential $\Phi_{\omega}$ we formulate and then solve the
system of equations for the dynamics of the induced charge
density $\tilde{\rho}_{\omega}=-Q_{\omega}\delta (z)$.
This dynamics is controlled by the transport along
the 2D layer, as well as by the nonzero current perpendicular
to the layer $j_n(\omega ,{\bf R})$.
The  conservation of charge  (in imaginary time) reads,
\begin{equation}
|\omega |Q_{\omega}-D_2\nabla^2_{{\bf R}}Q_{\omega}+\sigma_
2\nabla_{{\bf R}}^2\Phi_{\omega}=j_n,
\label{dynam1}
\end{equation}
where we introduced the 2D conductivity $\sigma_2({\bf B})=e^2\nu_
2D_2({\bf B})$,
and $\nu_2$ is the Fermi density of 2D states. 
On the left hand side of
this equation the
second and third terms are the divergences of the diffusion
and electrical currents respectively. 
The first two terms of this equation coincide with the left-hand side of 
Eq.\ (\ref{diffeq}) for the diffusion propagator $Q_{\omega}$.
This is precisely the reason for the cancellation of the diffusion 
pole discussed below.
Now, we can use this fact to eliminate $Q_{\omega}$ from the last equation:
\begin{equation}
j_n-\sigma_2\nabla_{{\bf R}}^2\Phi_{\omega}=-e\:sign (
\omega )\delta ({\bf R}-{\bf R}_0).
\label{dynam2}
\end{equation}

On the other hand, in the bulk of the conductor the charge is
not accumulated, $\tilde{\rho}_{\omega}|_{z>0}=0$. The conservation of 
charge then leads
to $\nabla {\bf j}(\omega ,{\bf r})=0$ and ${\bf n}\!\cdot\!{\bf j}
(\omega ,{\bf r})|_{z=0}=j_n(\omega ,{\bf R})$, where ${\bf j}$ is the 
density of current
in the bulk of the conductor. These two equations can be expressed
in terms of $\phi_{\omega}$:
\begin{equation}
\nabla_{{\bf r}}\!\cdot\!\hat{\sigma }\nabla_{{\bf r}}\phi_{
\omega}({\bf R},z)=0,
\label{dynam3}
\end{equation}
\begin{equation}
{\bf n}\!\cdot\!\hat{\sigma }\nabla_{
{\bf r}}\phi_{\omega}({\bf R},z)|_{z=0}=-j_n(\omega ,{\bf R}),
\label{dynam4}
\end{equation}
where $\hat{\sigma }\equiv\hat{\sigma }({\bf B})$ is the conductivity
tensor in the bulk of the conductor.
If the magnetic field is perpendicular to the surface of the conductor
($\varphi =0$), the conductivity tensor takes the simple form:
\begin{eqnarray}
\sigma_{xx}=\sigma_{yy}=
\frac{\sigma_3^{\scriptscriptstyle (0)}}{1+\omega_c^2\tau_3^2},\quad\sigma_{zz}=
\sigma_3^{\scriptscriptstyle (0)}\;,
\label{tensor1}\\
\sigma_{xy}=-\sigma_{yx}=-sign(\omega )\,{{\sigma_3^{\scriptscriptstyle (0)}
\omega_c\tau_
3}\over {1+\omega_c^2\tau_3^2}}\;,
\label{tensor2}
\end{eqnarray}
and all other elements vanish.
Here, $\sigma_3^{\scriptscriptstyle (0)}$ is the conductivity of 3D electrons in the case of zero
magnetic field, and $\tau_3$ is the momentum relaxation time of 3D
electrons.
In the case of arbitrary magnetic field orientation,
$\hat{\sigma}$ can be calculated by the rotation over the angle 
$\varphi$,
$\hat{\sigma}({\bf B})=\hat {U}(-\varphi )\hat{\sigma}({\bf B}
)|_{\varphi =0}\hat {U}(\varphi )$.

Solving Eqs.\ (\ref{dynam2},\ref{dynam3},\ref{dynam4}) simultaneously, we
get,
\begin{equation}
\Phi_{\omega}({\bf k})=-{{e\:sign( \omega)}
\over {\sigma_3^{\scriptscriptstyle (0)}
Z+\sigma_2{\bf k}^2}},
\label{Phi}
\end{equation}
where the function $Z({\bf B},{\bf k})$ is given by
\begin{equation}
Z({\bf B},{\bf k})=
\frac{\sqrt {k_x^2 \cos^2\alpha + k_y^2}+i k_x \sin \alpha}
{\sqrt {1+\omega_c^2\tau_3^2}},
\label{Z}
\end{equation}
with the angle $\alpha$ defined by
\begin{equation}
\frac{\sin \alpha}{\sin \varphi} =
sign(\omega){{\omega_c\tau_3}\over {\sqrt {
1+\omega_c^2\tau_3^2}}}.
\label{alpha}
\end{equation}
Finally, we substitute $Q_{\omega}$ from (\ref{diffuson}) and $\Phi_{\omega}$
from (\ref{Phi})
into Eq.\ (\ref{action2}), and arrive at the following formula
for the action:
\begin{equation}
S(\omega)=-{{e^2}\over {8\pi^2}}\int{{d{\bf k}}
\over {\left(|\omega|+D_2{\bf k}^2\right)\left(\sigma_3^{\scriptscriptstyle (0)}
Z+\sigma_2{\bf k}^2\right)}}.
\label{action3}
\end{equation}
We would like to mention the cancellation of the diffusion pole
in the expressions (\ref{Phi}) and (\ref{action3}) 
(compare to Eqs.\ (\ref{phi},\ref{action})). 
This well known fact 
(see for example Ref.\ \onlinecite{Altshu85}) physically
means that after tunneling, the  accommodation of the charge 
is entirely governed by plasmon modes.

In the equation (\ref{action3}) two terms in the denominator of the integrand,
$\sigma_3^{\scriptscriptstyle (0)}Z$ and $\sigma_2{\bf k}^2$,
are contributions from the screening of the
tunneling electron by 3D and 2D electrons, respectively.
In principle, one can expect to observe the crossover
from 3D ZBA with $S(\omega)\sim 1/\sqrt {\omega}$ 
($g=\sigma_3^{\scriptscriptstyle (0)}
Z/\sigma_2{\bf k}^2\gg 1$)
to 2D ZBA with $S(\omega)\sim 1/\omega$
($g\ll 1$).
In our case,
however, the  screening by 2D electrons
is weak.
Indeed, the integral
(\ref{action3}) 
should be evaluated for $\omega\sim 1/\tau_2$.
For $\varphi =0$, 
the simple estimate then gives
 $g\sim k_F l_3\gg 1$
in the case of zero magnetic field, 
and $g\sim k_F l_2\gg 1$ 
in strong magnetic field, $\omega_c\tau_{2,3}\gg 1$ 
(here, $l_2$ is the mean free path in the 2D layer). 
For $\varphi =\pi/2$ we have $g\sim \varepsilon_F/\omega_c\gg 1$
in strong magnetic field.
Therefore, we can neglect screening
by 2D electrons.
This equally means that in Eq.\ (\ref{dynam1}), for the induced charge density
$Q_{\omega}$, we can neglect the 2D current, 
$-\sigma_2\nabla_{{\bf R}}^2\Phi_{\omega}$, compared to the 3D current, $j_n$
(given by Eq.\ (\ref{dynam4})).                  
Thus, after tunneling  
the charge relaxation process has 3D character.
This leads to a $\sqrt{V}$ dependence of the differential
conductance, as is usual for 3D. We show this next.

Omitting the term $\sigma_2{\bf k}^2$
in the denominator
in the right hand side of (\ref{action3}),
we carry out the integration over ${\bf k}$ and
obtain,
\begin{equation}
S(\omega)=-{{e^2}\over {4\pi\sigma_3^{\scriptscriptstyle (0)}
\sqrt {|\omega |D_2^{\scriptscriptstyle (0)}}}}F(B,\varphi ),
\label{action4}
\end{equation}
\begin{equation}
 F(B,\varphi )=\sqrt {\left(1+\omega_c^2\tau_3^2\right)\left(1+\omega_
c^2\tau_2^2\cos^2\varphi\right)}E(\sin \alpha ),
\label{F1}
\end{equation}
where $E(s)=\int\limits_0^{\pi /2}d\theta\sqrt {1-s^2\sin^2\theta}\;$
is the complete elliptic integral. Substituting the action $S(\omega)$
from Eq.\ (\ref{action4}) into Eq.\ (\ref{IV}), we arrive at the 
final result,
\begin{equation}
{{1}\over{G}}{{d G}\over {d V}}=
{{e^3}\over {2\pi^2\sigma_3^{\scriptscriptstyle (0)}
\sqrt {2eVD_2^{\scriptscriptstyle (0)}}}}F(B,\varphi ).
\label{IV2}
\end{equation}
The correction to the differential conductance then
takes the form $\delta G\sim \sqrt{V}$.

Next we concentrate on the magnetic field dependence of the ZBA.
The fact that the magnetic field and $V$ dependencies
of the differential conductance are completely factorized 
allows us to represent the normalized amplitude of the ZBA,
$A(B,\varphi )\equiv -\delta G/G|_{V=0}$,  in a simple form.
To do this, we integrate Eq.\ (\ref{IV2}) over
$V$ and cut the integral at $eV\sim\tau_2^{-1}$.
We then arrive at the following
result:
\begin{equation}
A(B,\varphi )=A_0 F(B,\varphi ),
\label{amplitude}
\end{equation}
where $A_0 = \kappa \lambda_F^2/l_2l_3$, 
and $\kappa$ is a dimensionless number of order 1.
The equation (\ref{amplitude}), together with 
Eq.\ (\ref{F1}), represents the general result which is valid for an
arbitrary
magnetic field.
Now we consider the most interesting case of strong magnetic
field,
$\omega_c\tau_3\gg 1$.
In this limit $\sin^2\alpha =\sin^2\varphi$ and introducing the 
dimensionless parameter $h=\omega_c\tau_2$ we can write,
\begin{equation}
A(B,\varphi )=\kappa\left(\lambda_F/l_2\right)^2h\sqrt {1+h^2\cos^
2\varphi}E(\sin \varphi ).
\label{amplitude2}
\end{equation}

We are now  in a position to compare the result of our theoretical
analysis with the experimental data.
When the magnetic field is
perpendicular to the interface of the tunnel junction, $\varphi =0$, we
have ($\omega_c\tau_2\gg 1$):
\begin{equation}
A(B,0)={{\pi\kappa}\over 2}\left({{\lambda_Fh}\over {l_2}}\right
)^2={{\pi\kappa}\over 2}\left({{\lambda_F}\over {R_c}}\right)^2,
\label{amplitude3}
\end{equation}
where $R_c$ is the cyclotron radius. The amplitude of the anomaly
thus goes as $B^2$.
As it is clearly seen from Figs.\ \ref{fig7} and \ref{fig9}, 
the experimental data for 
${\bf B} \parallel {\bf n}$ are close to this dependence for both structures. 
Remarkably, the amplitude of
the anomaly does not depend on $\tau_2$ and $\tau_3$ 
(see Eq.\ (\ref{amplitude3})).
However, contrary to one's first expectation, the ZBA cannot be
observed in a perfect 3D metal. Although the amplitude of the anomaly
stays constant with $\tau_{2,3}\to\infty$, the dip of the tunneling
conductance
gets narrower (its width is given by $\tau_2^{-1}$ )
 and finally shrinks.

When the  magnetic field is parallel to the junction interface, $\varphi =\pi
/2$, from Eq.\ (\ref{amplitude2}) we obtain, 
\begin{equation}
A(B,\pi /2)=\kappa\left({{\lambda_F}\over {l_2}}\right)^2h,
\label{amplitude4}
\end{equation}
i.e.\ the amplitude of ZBA is a linear function of 
magnetic field. The same dependence is observed experimentally 
(see Fig.\ \ref{fig9}).

Finally, we can keep the amplitude of the magnetic field constant
and study the angular dependence of the ZBA. From Eq.\
(\ref{amplitude2}) it follows that
\begin{equation}
{{A(B,\varphi )}\over {A(B,0)}}={2\over {\pi h}}\sqrt {1+h^2\cos^
2\varphi}E(\sin \varphi ).
\label{amplitude5}
\end{equation}
This dependence and the dependence corresponding to the 
pure 3D case \cite{Sukhor} are plotted in Fig.\ \ref{fig10}. 
One can see that the expression (\ref{amplitude5}) is in excellent 
agreement with the experimental data without any fitting parameters.

\section{Conclusion}

We have presented the results of experimental and theoretical studies of 
the zero-bias anomaly (ZBA) in tunnel structures with
2D and 3D electron states coexisting near the 
semiconductor surface. 
It has been shown that the specific scenario of tunneling realized in this 
structures is: (i) electrons tunnel mainly into 2D states, (ii) immediately 
after tunneling, the electrons move diffusively in a 2D layer, and (iii) the 
main contribution to screening comes from the 3D electrons and, as a result,
the charge relaxation has a 3D character. 

This leads to the 
peculiar features of the magnetic field dependence of the ZBA amplitude. 
When the magnetic field is perpendicular to the interface of the tunnel 
junction, the ZBA amplitude grows as $B^2$, 
in agreement with Ref.\ \onlinecite{Altshu79}.
Although the magnetic field 
dependence has strong anisotropy, as predicted in
Ref.\ \onlinecite{Sukhor}, it does not disappear completely when the magnetic 
field lies in the plane of the junction interface. 
Instead, the ZBA amplitude is linear in $B$.

The experimental data show that the ZBA amplitude oscillates 
with the magnetic field. The origin of the oscillations is the Landau 
quantization of electron states in the 2D layer.
The detailed investigation of this effect will be the subject of future
work.

\acknowledgments

We would like to thank J.\ Kyriakidis for a critical reading of the
manuscript and for very helpful discussion.
This work was supported in part by the RFBR through Grants  
97-02-16168 and 98-02-17286, the Russian Program 
{\it Physics of Solid State Nanostructures} through Grant  97-1091,
and the Program {\it University of Russia} through Grant 420
at the Institute of Physics and Applied Mathematics,
and by the Swiss National Science Foundation at the 
University of Basel (E.V.S.)


\begin{references}
\bibitem[\dagger]{Minkov} email: Grigori.Minkov@usu.ru
\bibitem[*]{Eugene} email: sukhorukov@ubaclu.unibas.ch
\bibitem[\ddagger]{inst} On leave from
Institute of Microelectronics Technology,
Russian Academy of Sciences,
Chernogolovka, 142432 Russia.
\bibitem{Altshu85} 
   B.\ L.\ Altshuler, and A.\ G.\ Aronov, in
   {\em Electron-Electron Interaction in Disordered Systems},
        edited by A.\ L.\ Efros, and M.\ Pollak
        (Elsevier, Amsterdam, 1985).
\bibitem{Wolf} 
   V.\ N.\ Lutskii, A.\ S.\ Rylik, and A.\ K.\ Savchenko,
   Pis'ma Zh.\ Exsp.\ Teor.\ Fiz.\ {\bf 41}, 134 (1985)
   [JETP Lett.\ {\bf 41}, 163 (1985)];
   Alice E.\ White, R.\ C.\ Dynes, and J.\ P.\ Garno,
   Phys.\ Rev.\ B{\bf 31}, 1174 (1985);
   M.\ E.\ Gershenzon, V.\ N.\ Gubankov, and M.\ I.\ Falei,
   Zh.\ Eksp.\ Teor.\ Fiz.\ {\bf 90}, 2196 (1986)   
   [Sov.\ Phys.\ JETP {\bf 63}, 1287 (1986)];
   J.\ M.\ Valles, Jr., R.\ C.\ Dynes, and J.\ P.\ Garno,
   Phys.\ Rev.\ B{\bf 40}, 7590 (1989);
   P.\ Delsing, K.\ K.\ Likharev, L.\ S.\ Kuzmin, and T.\ Claeson, 
   Phys.\ Rev.\ Lett.\ {\bf 63}, 1180 (1989);
   R.\ C.\ Ashoori, J.\ A.\ Lebens, N.\ P.\ Bigelov, and R.\ H.\ Silsbee,
   Phys.\ Rev.\ B{\bf 48}, 4616 (1993);
   Shih-Ying Hsu and J.\ M.\ Valles, Jr.,
   Phys.\ Rev.\ B{\bf 49}, 16600 (1994);
   J.\ P.\ Kauppinen and J.\ P.\ Pekola, 
   Phys.\ Rev.\ Lett.\ {\bf 77}, 3889 (1996); 
   D.\ N.\ Davidov, J.\ Haruyama, D.\ Routkevitch, B.\ V.\ Statt,
   M.\ Moskovits, and J.\ M.\ Xu,
   Phys.\ Rev.\ B{\bf 57}, 13550 (1998);
   T.\ A.\ Polyanskaya, T.\ Yu.\ Allen, Kh.\ G.\ Nazhmudinov, 
   and I.\ G.\ Savel'ev,
   Semiconductors, {\bf 32}, 517 (1998).
   For the review of early experiments, see
   E.\ L.\ Wolf, {\em Principles of Electron Tunneling Spectroscopy}
              (Clarendon Press, Oxford, 1985).
\bibitem{Lee} 
   B.\ L.\ Altshuler and A.\ G.\ Aronov, 
   Solid State Commun.\ {\bf 30}, 115 (1979); 
   B.\ L.\ Altshuler,  A.\ G.\ Aronov and P.\ A.\ Lee, 
   Phys.\ Rev.\ Lett.\ {\bf 44}, 1288 (1980).
\bibitem{Altshu79}
   B.\ L.\ Altshuler, and A.\ G.\ Aronov,
   Zh.\ Eksp.\ Teor.\ Fiz.\ {\bf 77}, 2028 (1979)   
   [Sov.\ Phys.\ JETP {\bf 50}, 968 (1979)].
\bibitem{Nazarov1} 
   Yu.\ V.\ Nazarov, Zh.\ Eksp.\ Teor.\ Fiz.\ {\bf 95}, 975 (1989)
   [Sov.\ Phys.\ JETP {\bf 68}, 561 (1989)].
\bibitem{Nazarov2} 
   Yu.\ V.\ Nazarov, Fiz.\ Tverd.\ Tela {\bf 31}, 188 (1989)
   [Sov.\ Phys.\ Solid State {\bf 31}, 1581 (1989)].
\bibitem{Devoret} 
   M.\ H.\ Devoret, D.\ Esteve, H.\ Grabert,
   G.-L.\ Ingold, H.\ Pothier, and C.\ Urbina,
   Phys.\ Rev.\ Lett.\ {\bf 64}, 1824 (1990).
\bibitem{Ingold} 
   G.-L.\ Ingold, and Yu.\ V.\ Nazarov, in   
   {\em Single Charge Tunneling},
   edited by H.\ Grabert and M.\ H.\ Devoret (Plenum, New York, 1992).
\bibitem{Lev} 
   L.\ S.\ Levitov, and A.\ V.\ Shytov,
   Pis'ma Zh.\ Eksp.\ Teor.\ Fiz.\ {\bf 66}, 200 (1997)
   [Sov.\ Phys.\ JETP Lett.\ {\bf 66}, 214 (1997)].
\bibitem{Altshu84}
   B.\ L.\ Altshuler, A.\ G.\ Aronov, and A.\ Yu.\ Zuzin,
   Zh.\ Eksp.\ Teor.\ Fiz.\ {\bf 86}, 709 (1984)   
   [Sov.\ Phys.\ JETP {\bf 59}, 415 (1984)].
\bibitem{Sukhor} 
   E.\ V.\ Sukhorukov and A.\ V.\ Khaetskii, 
   Phys.\ Rev.\ B{\bf 56}, 1456 (1997).
\bibitem{anisotropy} 
   I.\ N.\ Kotel'nikov, A.\ S.\ Rylik, and A.\ Ya.\ Shul'man,
   Pis'ma Zh.\ Eksp.\ Teor.\ Fiz.\ {\bf 58}, 831 (1993)
   [Sov.\ Phys.\ JETP Lett.\ {\bf 58}, 779 (1993)].
\bibitem{Dub}
   Yu.\ V.\ Dubrovskii, Yu.\ N.\ Khanin, T.\ G.\ Andersson, 
   U.\ Gennser, D.\ K.\ Maude, and J.-C.\ Portal,
   Zh.\ Eksp.\ Teor.\ Fiz.\ {\bf 109}, 868 (1996)
   [Sov.\ Phys.\ JETP {\bf 82}, 467 (1996)].
\bibitem{Tsui1} 
   D.\ C.\ Tsui, Phys.\ Rev.\ B{\bf 8}, 2657 (1973). 
\bibitem{Tsui2} 
   D.\ C.\ Tsui, Phys.\ Rev.\ B{\bf 12}, 5739 (1975).
\bibitem{Tsui3} 
   D.\ C.\ Tsui, Phys.\ Rev.\ B{\bf 12}, 5853 (1975).
\bibitem{Minkov95} 
   G.\ M.\ Minkov, A.\ V.\ Germanenko, V.\ A.\ Larionova, 
   O.\ E.\ Rut, Sem.\ Sci.\ Techn.\ {\bf 10}, 1578 (1995).
\bibitem{Minkov73} 
   G.\ M.\ Minkov, A.\ V.\ Germanenko, V.\ A.\ Larionova, 
   O.\ E.\ Rut, Phys.\ Rev.\ B{\bf 54}, 1841 (1996).
\bibitem{Bardin} 
   J.\ Bardin, Phys.\ Rev.\ Lett.\ {\bf 6}, 57 (1961).
\bibitem{rate23}
   Eric D.\ Siggia, and P.\ C.\ Kwok, Phys.\ Rev.\ B{\bf 2}, 1024 
   (1970).

\end{references}
\end{document}